\documentstyle[aps,preprint,epsfig]{revtex} 

\def\be{\begin{equation}} 							
\def\ee{\end{equation}} 
\def\bea{\begin{eqnarray}} 
\def\eea{\end{eqnarray}} 
\def\no{\nonumber} 
\def\oli{\overline} 

\tightenlines 
\begin{document} 
\draft

\title{Resonant Production of Scalar Diquarks at the Next Generation
Electron--Positron Colliders}

\author{ Andr\'e Gusso \footnote{e-mail:gusso@fisica.ufpr.br}}

\address{Instituto de F\'{\i}sica Te\'orica, 
         Universidade  Estadual Paulista, \\  
         Rua Pamplona 145, 01405--900 S\~ao Paulo -- SP, Brazil} 

\maketitle

\thispagestyle{empty}
 
\begin{abstract}
We investigate the potential of TESLA and JLC/NLC electron-positron
linear collider designs to observe diquarks produced resonantly in
processes involving hard photons. 
\end{abstract}

\section{Introduction}
\label{introduction}
 
Diquarks are amongst those particles whose features make them worth to
be studied carefully. Like leptoquarks and dileptons, those particles
are naturally predicted by many interesting extensions of the standard
model as well as are invoked in other models in order to solve some of
the most puzzling problems  found in particle physics. These
particles, also known as biquarks and duarks,  carry baryon number
$B=\pm2/3$ and lepton number $L = 0$, are colored and interact with
quarks but not with leptons, and  have integer spin.

Scalar diquarks are predicted in many Grand Unified Theories (with or
without supersymmetry) \cite{SIM} and models of composite particles
\cite{Wudka},  are invoked as an essential ingredient for a model that
evades the strong $CP$ problem \cite{cp} and  in \cite{kaon} a class
of models containing  scalar diquarks  is considered that can generate
large $CP$ conserving and violating contributions  to double Cabibbo
suppressed decays without affecting $D^0 -\oli{D^0}$ mixing. Vector
diquarks are predicted by models of composite particles \cite{Wudka}. 

Diquark production was already analyzed for $e^+ e^-$, $e p$,
$p\oli{p}$ and $pp$  \cite{ee,ep,ppb,pp} colliders. They are
expected to be easily observable at LHC \cite{pp} through their
resonant production followed by the decay to two jets. Scalar and
vector diquarks with masses up to approximately 10 TeV could be
observed, depending on their couplings to the quarks. If relatively
light diquarks are observed they can also be expected to be on the
reach of the next generation linear colliders, whose center of mass
energy ranges from 500 GeV up to 1 TeV in the most conservative
designs, like TESLA \cite{Tesla} and JLC/NLC \cite{nlc}. The best
channel for their observation on these machines would be, in
principle, the pair production of diquarks. However, this channel is
only useful if the diquark masses are smaller than $\sqrt{s}/2$. This
is a great limitation {\it per se}, and becomes more  problematic when
the present experimental limits on $E_6$  superstring inspired
model scalar diquarks mass is taken into account, $290 <
M_{\phi_{E_6}} < 420$ GeV \cite{limits}. 

\begin{figure}
\centerline{\mbox{\epsfig{file=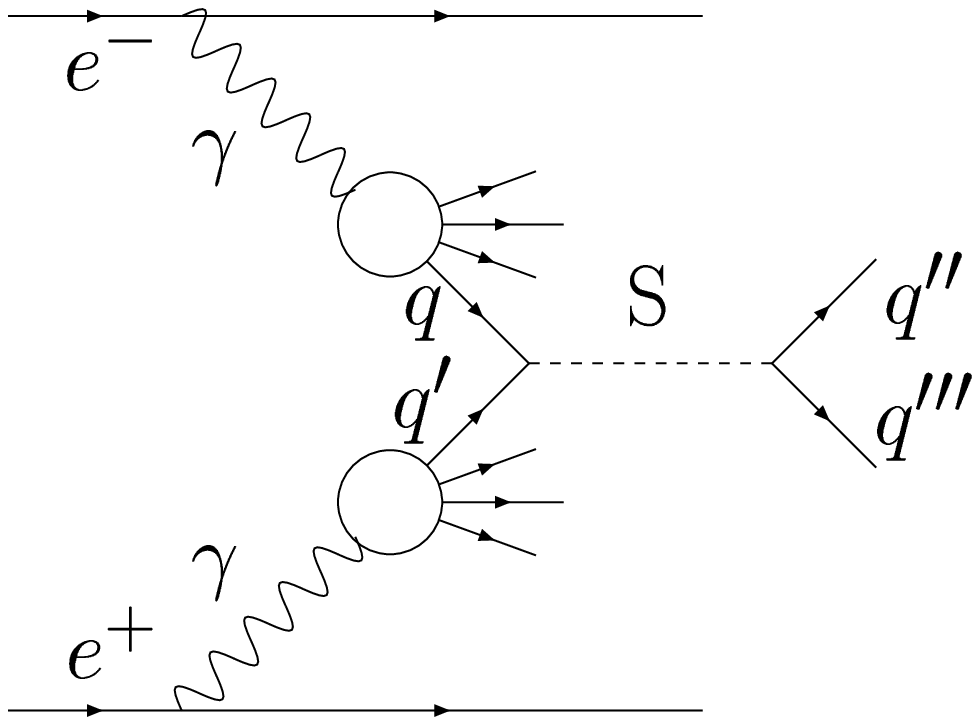,width=0.7\textwidth}}}
\caption{Schematic representation of the process leading to the
resonant production of diquarks in $e^+ e^-$ colliders.}
\label{resofig}
\end{figure}

In order to overcome this limitation we study the possibility of
observing diquarks produced resonantly through the process depicted in
Fig. \ref{resofig}. The resonant production of new particles through
higher-order processes has already shown its usefulness as shown, for
example, in \cite{dileptons} for the resonant production o dileptons.
In this article we will focus on the production of scalar diquarks.

This article is organized as follows. In Sect. \ref{signal} we
introduce the effective Lagrangians relevant to our analysis, and
other theoretical  and computational tools required to determine the
cross-section for the process depicted in Fig. \ref{resofig}. In Sect.
\ref{background} we analyze the relevant background. In Sect.
\ref{results} the results for the significance of the signal over
background are presented. Finally, we conclude in Sect.
\ref{conclusion}.

\section{Signal}
\label{signal}

Each extension of the standard model containing scalar diquarks we
referred to in the Introduction  has its won Lagragangian  coupling
diquarks with  quarks. Yet, other models can be created and contain
other Lagrangians. To take into account both the existing models and
the ones that can be devised in the future we consider the most
general $SU(3)_c \times SU(2)_L \times U(1)_Y$ invariant Lagrangian
coupling scalar diquarks with the quarks of the standard model. The
coupling of diquarks to other bosons and with itself are of no
relevance here. Following Ref. \cite{Nieves}, the possible scalar
diquarks are $SU(3)$ triplets and sextets (In what follows, $i,j$ and
$k$  are $SU(3)$ indices while $\alpha$ and $\beta$ are $SU(2)$, and
family indices are omitted)
\bea
S^{i} (3,1,-1/3), \,\, &S'^{\, i} (3,1,-4/3),& \,\, S''^{\, i} (3,3,-1/3)
, \,\, S'''^{i} (3,1,2/3), \,\, \no \\
S^{ij} (6,1,1/3), \,\, &S'^{\, ij} (6,3,1/3),& \,\, S''^{\, ij} (6,1,4/3)
, \,\, S'''^{\, ij} (6,1,-2/3)  ,
\eea
where the $SU(3)_c \times SU(2)_L \times U(1)_Y$ multiplicities are
given in parenthesis. Their couplings to the quark singlets $u$ and
$d$, and doublets $q$ are given by
\bea
{\cal L}_S &=&  \lambda_1 S^{i}  \epsilon_{ijk} \epsilon_{\alpha \beta} \, 
\oli{q_R^c}^{\alpha j} q_L^{\beta k} + 
 \lambda_2 S^{i}  \epsilon_{ijk} \, \oli{u_L^c}^j  d_R^{k} + 
 \lambda_3 S'^{\, i}  \epsilon_{ijk} \, \oli{u_L^c}^j  u_R^{k} \no \\
 &&+ \lambda_4 S''^{\, i}  \epsilon_{ijk} \, \oli{q_R^c}^{\alpha j}
 ({\bf \epsilon \tau})_{\alpha \beta} \, q_L^{\beta k} +
\lambda_5 S'''^{\, i}  \epsilon_{ijk} \, \oli{d_L^c}^j  d_R^{k} +
  \lambda_6 S^{ij} \epsilon^{\alpha \beta} \, \oli{q}_{L \alpha i} 
  q^c_{R \beta j} \no \\
  &&+  \lambda_7 S^{ij}  \oli{d}_{R i} u^c_{L j} + 
\lambda_8 S'^{\, ij} \oli{q}_{L \alpha i}({\bf \tau \epsilon})^{\alpha 
\beta}
 q_{R \beta j}^c + \lambda_9 S''^{\, ij}  \oli{u}_{R i} u^c_{L j} + 
\lambda_{10} S'''^{\, ij}  \oli{d}_{R i} d^c_{L j} . \no \\
\label{laesc}
\eea

From (\ref{laesc}) we can get the Feynman rules required to  determine
the cross-section for the process in Fig. \ref{resofig}. In a process
like this, most of the particles in the final state  are expected to
be lost in the beam pipe, the two exceptions being the quarks
resulting from diquark decay. Consequently, the relevant signal is a
$p_T$ balanced coplanar  pair of jets. 

The cross-section for the production of a pair of jets through the
process in Fig. \ref{resofig} is given to good accuracy by 
\be
\sigma(\hat{s})_{e^+ e^- \rightarrow 2 \; jets} =  
\int_{x_{min}}^1 dx  
\int_{y_{min}}^1 dy f_{pe}(x)  f_{p'e}(y) 
 \sigma(\hat{s})_{q q' \rightarrow q''  q'''} \delta(x y s - \hat{s})
\label{xsection}
\ee
In this expression, $f_{pe}(x)$ corresponds to the probability density
of finding a parton $p$ carrying a fraction $x$ of the energy of the
electrons (or positrons) in the initial beams.  $\sigma(\hat{s})_{q q'
\rightarrow q''  q'''}$ corresponds to the cross-section for the
subprocess in which two on-shell quarks  $q$ and $q'$, with total
energy $\sqrt{\hat{s}} = \sqrt{xys}$, interact resulting into a diquark that
utterly decays to the quarks $q''$ and $q'''$. The lower limits of
integration $x_{min}$ and $y_{min}$, guarantees the energy required for
the production of quarks. Because we are going to consider the five
light quarks as massless those limits are actually equal to zero.

\begin{figure}
\centerline{\mbox{\epsfig{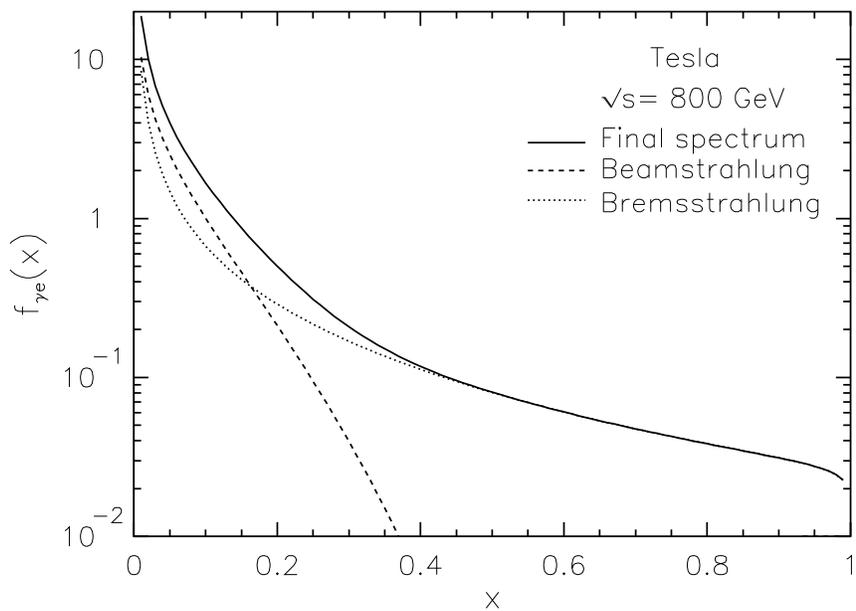}}}
\caption{The probability density $f_{\gamma e}$ for TESLA operating at
$\sqrt{s} = 800$ GeV.}
\label{fge_fig}
\end{figure}

The probability $f_{pe}(x)$ results from the convolution of the 
spectrum of photons emitted by the initial electrons, $f_{\gamma
e}(x)$, with $f_{p\gamma}(x)$ the photonic parton distribution 
function (PDF)
\be
f_{pe}(x) = \int_{0}^1 dz  \int_{0}^1 dy f_{\gamma e}(z) f_{p\gamma}(y)
\delta(z y - x)  = \int_x^1 \frac{dz}{z} f_{\gamma e}(z) f_{p\gamma}
\left(\frac{x}{z} \right) .
\ee
The photon spectrum  $f_{\gamma e}$ results from the combination of
two effects, namely,  bremsstrahlung and beamstrahlung. The photon
spectrum resulting from bremsstrahlung for an electron beam with energy
$\sqrt{s}/2$ is \cite{deFlorian:1999ge}
\be
 f_{\gamma e}^{brems}(x)=\frac{\alpha_{\rm em}}{2\pi}\left[
 \frac{1+(1-x)^2}{x}\ln\frac{s(1-x)^2}{m_e^2 x^2}
 +2 m_e^2 x\left(\frac{1}{s(1-x)}-\frac{1-x}{m_e^2 x^2}\right)\right] \,  .
\label{brems}
\ee
The photon spectrum that results from the electromagnetic interaction
 between the electron and positron beams  in the intersection region,
 is described by a more complicated expression \cite{Chen:1992wd}
\bea
\label{beam}
 f_{\gamma e}^{beam}(x) &=& \frac{1}{\Gamma\left(\frac{1}{3}\right)}
  \left(\frac{2}{3\Upsilon}\right)^\frac{1}{3} x^{-\frac{2}{3}}
  (1-x)^{-\frac{1}{3}} \exp\left[-\frac{2 x}{3\Upsilon (1-x)}\right] \no \\
 &&\times  \biggl\{\frac{1-\sqrt{\frac{\Upsilon}{24}}}{g(x)}\left[ 1-\frac{1}
  {g(x) N_\gamma}\left(1-e^{-g(x) N_\gamma}\right)\right] \no \\
  &&+ \sqrt{\frac{\Upsilon}{24}} \left[1-\frac{1}{N_\gamma} \left( 
  1-e^{-N_\gamma}\right)\right]\biggr\}\, ,
  \label{beamphoton}
\eea
The  beamstrahlung parameter $\Upsilon$ that characterizes each linear
collider is given by,
\be
 \Upsilon = \frac{5 r_e^2 E_e N}{6 \alpha \sigma_z (\sigma_x+\sigma_y) 
 m_e},
\ee
with $E_e= \sqrt{s}/2$, $r_e = \alpha_{\rm em}/m_e = 2.818 \cdot
10^{-15}$ the classical  electron radius, $\sigma_x,~\sigma_y$ and
$\sigma_z$ the average size of the particle bunches, and  $N$ the
number of particles per bunch. Other definitions in (\ref{beamphoton})
are
\be
 g(x) = 1 - \frac{1}{2} \left[(1+x)\sqrt{1+\Upsilon^{\frac{2}{3}}}+1-x\right]
 (1-x)^{\frac{2}{3}} \, , 
\ee
and the average number of  photons per electron irradiated during the
electron and positron bunches interaction
\be
 N_\gamma = \frac{5 \alpha^2 \sigma_z m_e}{2 r_e E_e} \frac{\Upsilon}
 {\sqrt{1+\Upsilon^{\frac{2}{3}}}} \, .
\ee
Because the  photon production by bremsstrahlung and beamstrahlung are
distinct physical processes the final spectrum is given simply by
$f_{\gamma e}(x) = f^{brems}_{\gamma e}(x) + f^{beam}_{\gamma e}(x)$.
In Fig. \ref{fge_fig} we present  $f_{\gamma e}(x)$ for TESLA
operating at $\sqrt{s} = 800$ GeV. For both TESLA and JLC/NLC at 
$\sqrt{s} = 500$ GeV  we obtain a similar spectrum, however for  
JLC/NLC at 1 TeV,  the beamstrahlung is significant up to $x \sim 0.7$
instead.

\begin{figure}
\centerline{\mbox{\epsfig{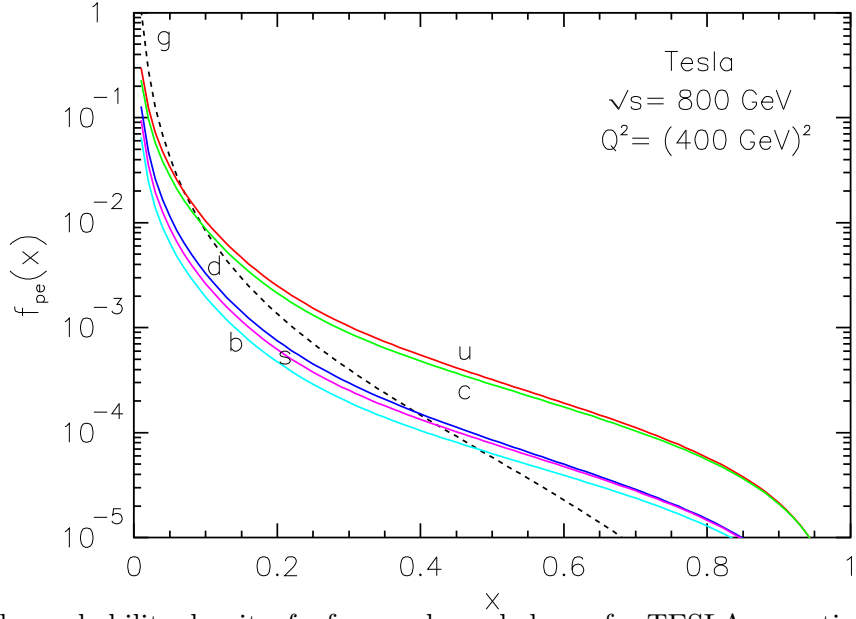}}}
\caption{The probability density $f_{pe}$ for quarks and gluons for
TESLA operating at $\sqrt{s} = 800$ GeV and an energy scale $Q^2 = (400
{\rm GeV})^2$.}
\label{fpe_fig}
\end{figure}

The photonic PDFs were determined by several groups each using
distinct physical models and different sets of experimental data
\cite{PDFs}. We adopted for our analysis the GRV photonic PDFs 
\cite{GRV}, calculated at leading order.  This choice is based on two
facts: (i) according to \cite{PDFs} the GRV parameterization  gives
one of the best descriptions of the structure function $F_2^\gamma$
and (ii) this parameterization is valid up to a characteristic scale
$Q^2 = 10^6$ GeV$^2$. Considering the characteristic scale for the
scattering processes equal to $\hat{s}/4$ implies that it is valid up
to $\sqrt{s} \sim 2$ TeV.  Recently, another  parameterization
\cite{CJKL} was developed, based on an improved model and more data,
which fits the data on $F_2^\gamma$ better than GRV parameterization.
However, both parameterizations fits the data almost equally well and
the parameterization in \cite{CJKL} presents the limitation of being
valid only up to $\sqrt{s} \sim 900$ GeV, while we are interested in 
energies up to 1 TeV. It is also  worth  mentioning that for the
relatively high values of $Q^2$ (the characteristic scale at which the
PDFs are evaluated) and $x$ (the energy fraction carried by the
parton) that will be important in our search for heavy diquarks, the
various  parameterizations predict quite similar parton distributions
\cite{CJKL}. In Fig. \ref{fpe_fig} we show $f_{pe}(x)$ for quarks and
gluons as expected for TESLA operating at $\sqrt{s} = 800$ GeV, for
$Q^2 = (400 {\rm GeV})^2$. Similar results hold for TESLA and
JLC/NLC.  

The last ingredient present in (\ref{xsection}) is the cross-section  
$\sigma(\hat{s})_{q q' \rightarrow  q''  q'''}$. In order to obtain
the correct contribution from the subprocess $q q' \rightarrow  q'' 
q'''$ we have to consider the interference of the $s$-channel 
production of the diquark with the  diagrams involving the gluon
exchange in both $t$ and $u$ channels. Denoting these contributions as
$a, b$ and $c$, respectively, the differential cross-section for the
signal generated by a scalar diquark of mass $M_S$ coupling to
quarks with strength $\lambda$  can be written as
\bea
\frac{d \sigma}{d t} &=& \frac{d \sigma(a)}{d t}
+ \frac{d \sigma(a \times b)}{d t} +  \frac{d \sigma(a \times c)}{d t} \no 
\\ &=&\frac{f_s \lambda´^4 }{16 \pi s [ (s-M_S^2)^2 + M_S^2\Gamma^2 ] }
 + \frac{f'_s \lambda´^2 \alpha_s  (s-M_S^2) }{ t [ (s-M_S^2)^2 + M_S^2 
 \Gamma^2 ] } + \frac{f''_s \lambda´^2 \alpha_s  (s-M_S^2) }{ u [
  (s-M_S^2)^2 + M_S^2 \Gamma^2 ] } . \no \\
\label{dsdt}
\eea
The decay width is 
\be
\Gamma = \frac{ F_s \lambda´^2 M_S}{16 \pi} .
\ee
The factors $f_s, f'_s, f''_s$ and $F_s$ depend upon  which of the
interactions presented in (\ref{laesc}) is considered and incorporate
statistical factors associated with the presence of identical fermions
in the vertices \cite{feynrules}. The sum and average over colors are incorporated in
those factors as well. $F_s$ contains the sums over all possible final
states for diquark decay. The cross-section (\ref{dsdt}) was derived
assuming massless quarks, a fairly good assumption at the energies we
are going to consider.

At this point we have the freedom to choose which of the interactions
from ${\cal L}_s$ are worth to be analyzed. The idea is to choose a
representative interaction whose results can be lately interpreted
approximately in terms of  the other interactions. For that aim it is
helpful to observe in  Fig. \ref{fpe_fig}  that $f_{pe}$ for $u$ and
$c$ quarks are at least three times bigger than for the other quarks.
Because the cross-section (\ref{xsection}) involves the product of two
$f_{pe}$, we are led to  conclude  that processes involving $u$ and
$c$ quarks are about ten times more likely to occur. For that reason
we choose the term   $\lambda'_9 S''^{\, ij}  \oli{u}_{R i} u^c_{L j}$
from ${\cal L}_S$, which involves $u$ and $c$ quarks. We also consider
$\lambda'_9$, that must be symmetric, diagonal in the flavor space.
Consequently the initial and final states in the subprocess in which
the diquark is produced  involve two $u$ or two $c$ quarks. For
comparison, if $\lambda'_9$ where flavor democratic, we could have 12
instead of 4 different subprocesses. The factors   $f_s, f'_s, f''_s$
and $F_s$ are 8, 8/9, 8/9 and  4, respectively, for the interaction we
have chosen, for both $u$ and $c$ quarks in the vertices.

\section{background}
\label{background}

The two-jet background receives contributions from the annihilation
process $e^+ e^-  \rightarrow \gamma, \, Z \rightarrow q \oli{q}$ and
the hard two photon process $e^+ e^- \rightarrow e^+ e^- q \oli{q}$
\cite{drees}.  The two photon processes can be of type direct, once
resolved and twice resolved. The last two processes, despite being
higher order in $\alpha_s$, contribute significantly at lower energies. 

The cross-section for the background generated by the annihilation
process is given by
\be
\sigma(\hat{s})_{annih} = \sum_q \int_0^1 dx \int_0^1 dy f_{ee}(x) f_{ee}(y) 
\sigma (\hat{s})_{e^+ e^- \rightarrow q \bar{q}} \, \delta(x y s - \hat{s}) ,
\ee
where  $f_{ee}(x)$ is the electron (and positron) spectrum, the
probability density that we get an electron with  a fraction $x$ of
the initial electron energy in the beams, and the sum runs over all
possible quarks in the final state. $\sigma (\hat{s})_{e^+ e^-
\rightarrow q \bar{q}}$ corresponds to the cross-section for the
process $e^+ e^- \rightarrow \gamma, Z \rightarrow q\bar{q}$.
$f_{ee}$  results from the  convolution of the   beamstrahlung and the
initial state  radiation (ISR) spectra
\be
f_{ee}(x) = \int_x^1 \frac{dz}{z} f_{ee}^{ISR}(z)
f^{beam}_{ee} \left( \frac{x}{z} \right)  .
\label{feeeq}
\ee
We use the following  analytical expression for $f_{ee}^{ISR}(x)$
\cite{fee},
\be
f_{ee}^{ISR}(x) = \frac{\beta}{2} (1-x)^{\frac{\beta}{2}-1}\left( 1+\frac38 
\beta \right) -\frac14 \beta (1+x)  ,
\ee
where $\beta$ is defined as,
\be
\beta = \frac{2 \alpha_{\rm em}}{\pi} \left[ \ln \left( \frac{s}{m_e^2} \right)
 -1 \right] .
\ee
The spectrum $f_{ee}^{beam}$ was evaluated numerically using CIRCE 
\cite{circe} a  FORTRAN routine containing parameterizations for
electron, positron and photon  spectra resulting from beamstrahlung
at the future $e^+ e^-$ colliders.

The cross-section for the background generated by two hard photon
processes is calculated from 
\be
\sigma(\hat{s})_{2 \, \gamma} = \sum_{i j} \sum_p \int_0^1 dx \int_0^1 dy 
\, f_{ie}(x) f_{je}(y) 
\sigma (\hat{s})_{i j  \rightarrow 2 jets} \, \delta(x y s - \hat{s}) ,
\ee
where the indexes $i, j = \gamma$ or $p$. The sum over $i$ and $j$
corresponds to the sum over direct, once and twice resolved processes
and the sum over $p$ remind us that we have to sum over subprocesses
involving different initial and final partons. In total we have to
consider one direct process, and 2 once resolved, and 8 twice resolved
subprocesses. The cross-sections for all such processes can be found
in \cite{Predazzi}. The five lighter quarks are assumed massless
throughout our analysis and the top quark, of course, is not
considered as a source of background.

\section{results}
\label{results}

In the previous sections we presented all the theoretical and
computational tools required for our analysis.  We used them to
determine the expected significance of the signal over  background
for collider parameters as given in \cite{Tesla}, for TESLA operating
at 500 GeV and 800 GeV, and  as given in \cite{nlcdesign} for NLC
operating at 500 GeV and 1 TeV. Because the designs for JLC and NLC
tends to converge, specially with respect to the parameters relevant
for the beamstrahlung effect, we can  in fact assume that our results
are valid for JLC/NLC design. We show in Table \ref{param_tab} the 
relevant parameters for the calculation of beamstrahlung.

The significance of signal over background is defined as,
\be
\sigma_S = \frac{S}{\sqrt{B}},
\label{sign}
\ee
where $S$ is the  number of expected two-jet events attributed to the
resonant production of diquarks, and $B$ is the  number of two-jet
events resulting from the standard model alone.  The signal 
generated by a diquark of mass $M_S$ and decay rate $\Gamma$ is
calculated integrating the cross-section in  the two-jet invariant
mass ($M$)  interval  $M_S - \Gamma < M < M_S + \Gamma$, which
embraces approximately 95 $\%$ of the events around the resonance. 
For a realistic analysis of the contribution of the background we
considered the energy resolution of the hadronic calorimeter. For
TESLA detector  it reads,
\be
\frac{\delta E}{E} \leq \frac{0.5}{\sqrt{E}}  + 0.04 .
\ee
The same energy resolution was assumed for JLC/NLC. The corresponding
two-jet invariant mass resolution is given approximately (we assume it
to be exact in our calculations) by $\delta M = 0.5 \sqrt{M} + 0.02828
M$.  The background B is calculated by integrating the cross-section
in the range   $M_S - \Delta M < M < M_S + \Delta M$, with $\Delta M =
{\rm max}(\Gamma, \delta M)$.

The significance (\ref{sign}), scales with the collider luminosity as
$\sqrt{\cal L}$. Consequently, high luminosities are desirable for a
good signal over background discrimination. Presently, TESLA is
expected to deliver the highest luminosities, 500 fb$^{-1}$/year,
while the luminosity for JLC/NLC is  planned to be in the range
100-200  fb$^{-1}$/year.  With those luminosities in mind we determined
the significance for the following total integrated luminosities:
${\cal L} = 100, 300$ and 500 fb$^{-1}$ for both collider designs and
additionally  ${\cal L} = 2000$ fb$^{-1}$ for TESLA. 

In determining $\sigma_S$ we considered some cuts on the
pseudo-rapidity $\eta$ and the transverse momentum $p_T$ of the jets.
We worked  $\sigma_S$ out for $|\eta| < 1$ and  2, and $|p_T| > 15$
GeV, for both jets. This cut on $p_T$ is about the minimum required to
guarantee that the assumption of massless quarks is valid for all
quarks, including the bottom quark, because it implies an adequate 
minimum energy for the jets. It also guarantees the validity of the
PDF parameterization and eliminates considerable amount of background.
As we discuss in more detail later, higher $p_T$ cuts did not improve
$\sigma_S$ significantly, while reducing the absolute number of signal
events to undesirably  low values. Our analysis is also limited to the
search for diquarks with masses above 100 GeV. Lighter  Diquarks could
in principle be pair produced by LEP, which reached more than 200 GeV
as its center of mass energy. Here we are interested in what LEP could
not reach. 

In Fig. \ref{tesla500} we present $S/\sqrt{B}$ as a function of the
diquark mass, $M_S$, for TESLA operating at 500 GeV. The solid and
dotted lines are for  $|\eta| < 1$ and 2, respectively. The eight
upper curves, both solid and dotted, correspond to a coupling with
strength given by $\alpha = \lambda^2/4 \pi = 0.1$, while the eight
lower curves are for $\alpha = 0.01$. For each $\alpha$ and cut on
$\eta$ there are four curves corresponding to the integrated
luminosities ${\cal L} = 100, 300, 500$ and 2000 fb$^{-1}$. The
curves for $\alpha = 0.01$ where scaled by a factor 0.1 in order to
clearly separate them from the curves for $\alpha = 0.1$. The
straight dashed lines correspond to $S/\sqrt{B} = 3$, the minimum
requirement for signal observability. The choice $\alpha = 0.1$
corresponds approximately to the maximum acceptable value to keep
higher order contributions under control. The coupling $\alpha =
0.01$ is approximately the minimum value  for which the signal
significance is worth to be analyzed.  For smaller values of $\alpha$
the signal significance is much smaller than 3 for masses greater
than $\sqrt{s}/2$. Fig. \ref{tesla800} shows the same curves for
TESLA operating at 800 GeV. The results for JLC/NLC operating at 500
GeV and 1 TeV are presented in Figs. \ref{nlc500} and \ref{nlc1000}
for ${\cal L} = 100, 300$ and 500 fb$^{-1}$.

\section{Final discussion and conclusion}
\label{conclusion}

As already mentioned at the Introduction the Tevatron collider has set
limits on the masses of a class of  scalar diquarks, $290 <
M_{\phi_{E_6}} < 420$ GeV which are expected to be approximately valid
for other scalar diquarks. There also exist indirect bounds imposed
on   Yukawa coupling constants for quarks and diquarks. Stringent
indirect bounds for couplings involving the second and third
generations where set from the limits on branching ratio for the
inclusive decay $B  \rightarrow X_s \gamma$ \cite{bsgbounds}. Looser
bounds on couplings involving other generations come from the analysis
of precision LEP data \cite{LEPbounds}, neutron-antineutron
oscillation \cite{nnbounds} and rare nucleon and meson decay
\cite{decaybounds}. The looser bounds generally permit couplings of
the order one as required for $\alpha = 0.1$ ($\lambda \sim 1.25$) the
higher value we admitted in our analysis.

\begin{figure}
\centerline{\mbox{\epsfig{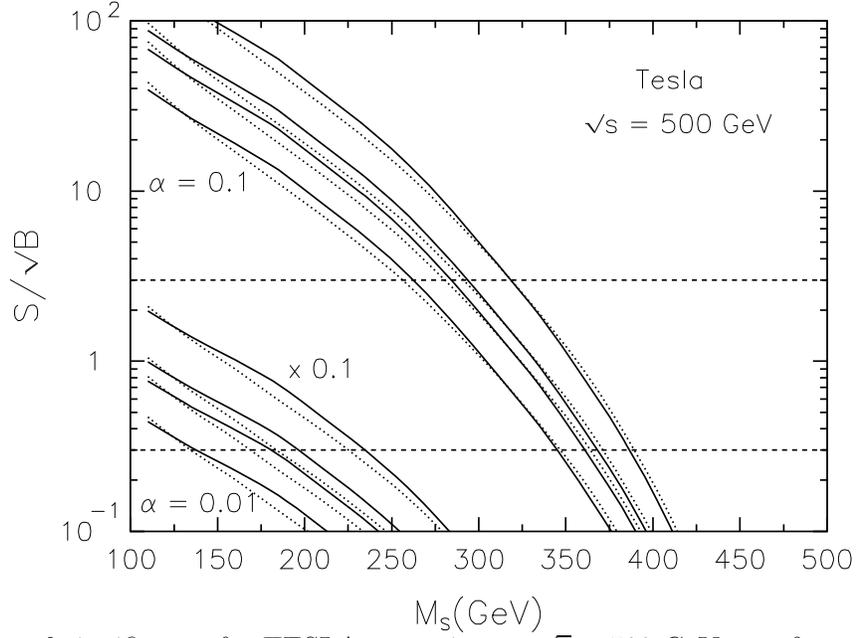}}}
\caption{Signal significance for TESLA operating at $\sqrt{s} = 500$
GeV as a function of diquark mass, $M_S$. Solid and dotted lines
correspond to $|\eta| < 1$ and 2, respectively. For all curves $|p_T|
> 15$ GeV.}
\label{tesla500}
\end{figure}

\begin{figure}
\centerline{\mbox{\epsfig{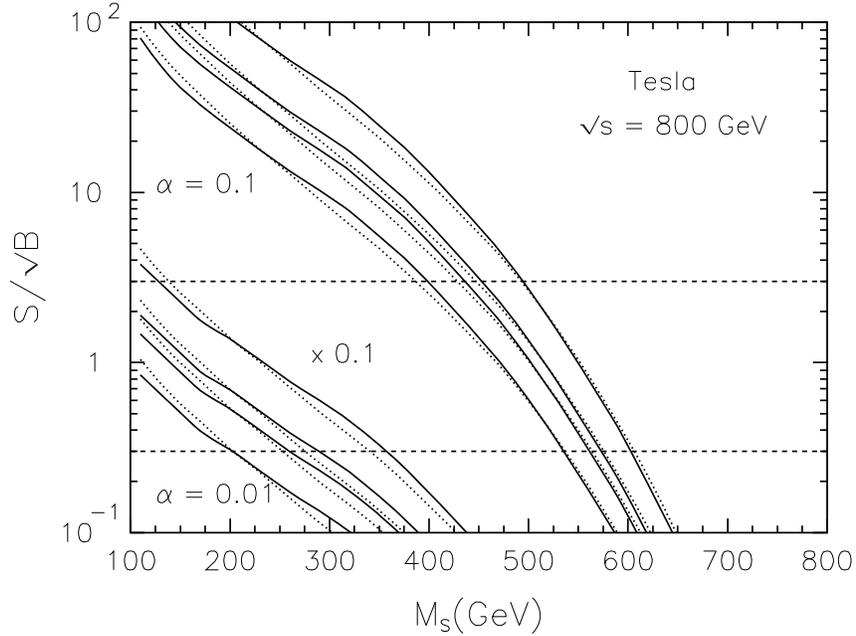}}}
\caption{The same as Fig. \ref{tesla500} for TESLA operating at
$\sqrt{s} = 800$ GeV.} 
\label{tesla800}
\end{figure}

\begin{figure}
\centerline{\mbox{\epsfig{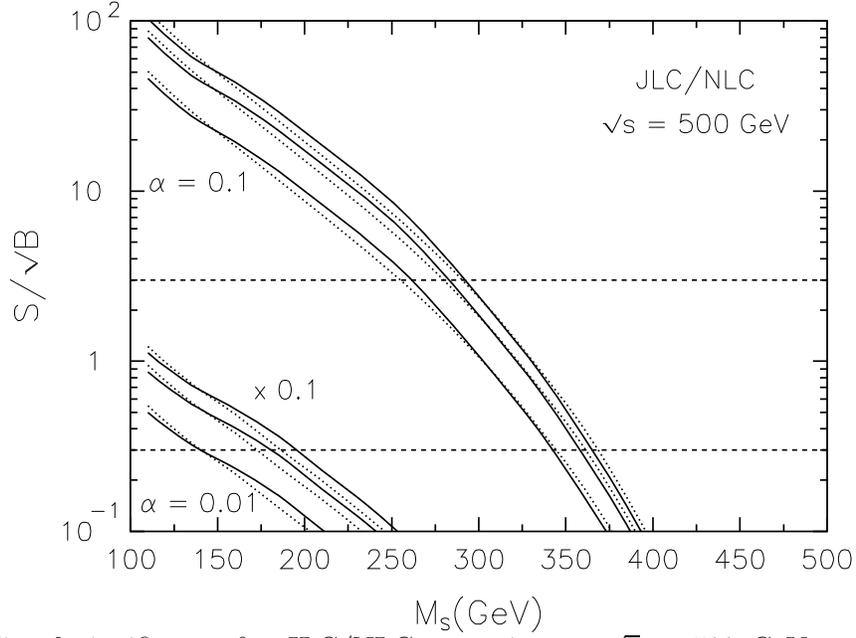}}}
\caption{Signal significance for JLC/NLC operating at $\sqrt{s} = 500$
GeV as a function of diquark mass, $M_S$. Solid and dotted lines
correspond to $|\eta| < 1$ and 2, respectively. For all curves $|p_T|
> 15$ GeV.}
\label{nlc500}
\end{figure}

\begin{figure}
\centerline{\mbox{\epsfig{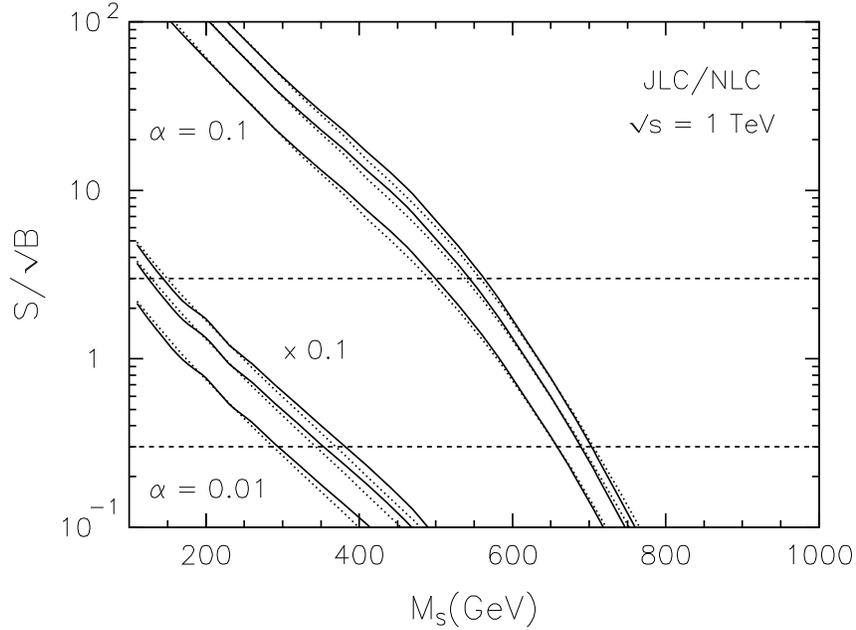}}}
\caption{The same as Fig. \ref{nlc500} for JLC/NLC operating at
$\sqrt{s} = 1$ TeV.} 
\label{nlc1000}
\end{figure}

It is exactly for relatively high values of the Yukawa coupling
constant that we may expect the resonant production of diquarks to be
useful. This conclusion can be draught from the curves for $\alpha =
0.1$ in Figs. \ref{tesla500}, \ref{tesla800}, \ref{nlc500} and
\ref{nlc1000}. For couplings smaller by just one order of magnitude
($\alpha = 0.01$) the resonant production does not produce any
interesting result. Is is worth mentioning at this point that for
$\alpha = 0.1$ the number of signal events is always of order of a few
tens or more for $\sigma_S > 3$, resulting in good statistics. Another
conclusion that can be draught is that the best signal over background
relation in the parameter region satisfying $\sigma_S > 3$ and $M_S >
\sqrt{s}/2$  is obtained considering the cut $|\eta|<1$.  However we 
can note from the results for the signal significance that the curves
for  $|\eta|<1$ are not always  above the corresponding curves for
$|\eta|<2$. This behavior for $\sigma_s$ can be attributed to the
distinct angular distributions for the background generated through
annihilation and for the background generated  by two hard photon
processes. The two hard photon processes prevail at energies below
$\sqrt{s}/2$ while the annihilation prevails at higher energies. This
partially explains the behavior of the curves. An additional aspect of
the background has to be taken into account,  the fact that once and
twice resolved processes initiated by gluons, that alter significantly
the angular  distribution, are more important at lower energies.
Altogether, we have the elements that imply the relatively complex
behavior of the curves for signal significance.

A final comment on the $p_T$ cuts. We have tested cuts on $p_T$
higher than 15 GeV, however they did not improve the signal over
background relation significantly. This can be explained by the fact
that the dependence on $p_T$ for the signal is almost the same as the
one for the background generated by hard two photon processes and  the
background from annihilation is almost constant with respect to
$p_T$.

We conclude, considering the present analysis and the previous works on
the search for diquarks at the next generation of linear colliders,
that diquarks with masses above $\sqrt{s}/2$ are unlikely to be seen
at such machines, unless new production channels are devised and prove
to be sufficiently sensitive to the presence of those particles. The
only exception are diquarks with relatively strong Yukawa coupling 
constants, that means $\lambda \sim 1$, with the additional
requirement of  coupling to $u$ and $c$ quarks.

\begin{acknowledgements}
This work was supported by Funda\c{c}\~ao de Amparo a Pesquisa do
Estado de S\~ao Paulo (FAPESP).
\end{acknowledgements}

\newpage


\begin{figure}
\centerline{\mbox{\epsfig{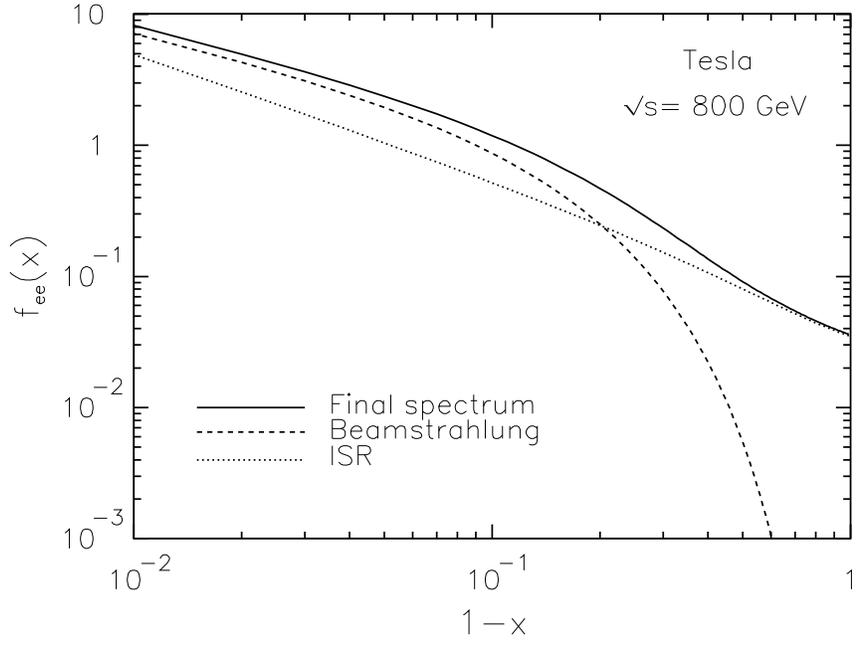}}}
\caption{The probability density $f_{ee}$ for TESLA operating at
$\sqrt{s} = 800$ GeV.}
\label{fee_fig}
\end{figure}

\begin{table}
\caption{Collider parameters relevant for the calculation of
beamstrahlung. $N$ is the number of particles per bunch,
$\sigma_{x,y,z}$ are the mean bunch dimensions ($z$ is the beam axis),
$\Upsilon$ is the beamstrahlung parameter and $N_\gamma$ the average
number of photons  per electron. }
\label{param_tab}
\begin{tabular}{ccccccc}
& Collider & TESLA  & TESLA & JLC/NLC & JLC/NLC \\ 
Parameter & &  500 GeV & 800 GeV & 500 GeV & 1 TeV \\ 
$N [10^{10}]$ & & 2.0 & 1.4 & 0.75 & 0.75 \\
$\sigma_x$ [nm] &  & 553 & 391 & 245 & 190 \\
$\sigma_y$ [nm] &  & 5 & 2.8 & 2.7 & 2.1 \\
$\sigma_z$ [$\mu$m] &  & 300 & 300 & 110 & 110 \\
$\Upsilon$ & & 0.053 & 0.084 & 0.122 & 0.315 \\
$N_\gamma$ & & 1.44 & 1.39 & 1.16 & 1.38 \\
\end{tabular}
\end{table}

\end{document}